\documentclass[10pt,conference]{IEEEtran}
\IEEEoverridecommandlockouts

\usepackage{cite}
\usepackage{amsmath,amssymb,amsfonts}
\usepackage{algorithmic}
\usepackage{graphicx}
\usepackage{textcomp}
\usepackage{xcolor}
\def\BibTeX{{\rm B\kern-.05em{\sc i\kern-.025em b}\kern-.08em
    T\kern-.1667em\lower.7ex\hbox{E}\kern-.125emX}}

\usepackage{hyperref}
\usepackage{enumitem}
\usepackage{booktabs}
\usepackage{siunitx}
\usepackage{pgfplots}
\usepackage{xfrac}
\pgfplotsset{compat=1.18}
\usepackage{subcaption}
\usepackage{balance}
\usepackage{soul}

\usetikzlibrary{shapes.geometric}

\newtheorem{definition}{Definition}

\newcommand*{\prob}{\operatorname{Pr}}

\newcommand*{\estimates}{\mathrel{\widehat=}}

\DeclareSIUnit\bpcu{bpcu}

\definecolor{plot0}{HTML}{004488}
\definecolor{plot1}{HTML}{DDAA33}
\definecolor{plot2}{HTML}{BB5566}
\definecolor{plot3}{HTML}{000000}
\definecolor{plot4}{HTML}{AAAAAA}

\begin{document}

\title{Modular Neural Wiretap Codes for Fading Channels
\thanks{This work was supported in part by the Federal Ministry of Education and Research of Germany (BMBF) within the project 6G-life, Project ID 16KISK001K, by the German Research Foundation (DFG) as part of Germany’s Excellence Strategy - EXC 2050/1 - Project ID 390696704 - Cluster of Excellence ``Centre for Tactile Internet with Human-in-the-Loop'' (CeTI) of Technische Universität Dresden, by the ZENITH Research and Leadership Career Development Fund, Chalmers Transport Area of Advance, and the ELLIIT funding endowed by the Swedish government.}
}

\author{\IEEEauthorblockN{Daniel Seifert\textsuperscript{*}, Onur Günlü\textsuperscript{\textdagger}, and Rafael F. Schaefer\textsuperscript{*}}
\IEEEauthorblockA{\textsuperscript{*}Chair of Information Theory and Machine Learning, 
Technische Universität Dresden, Germany \\
\textsuperscript{\textdagger}Information Theory and Security Laboratory (ITSL),
Linköping University, Sweden \\
\{daniel.seifert, rafael.schaefer\}@tu-dresden.de, onur.gunlu@liu.se}
}

\maketitle

\begin{abstract}
The wiretap channel is a well-studied problem in the physical layer security literature. Although it is proven that the decoding error probability and information leakage can be made arbitrarily small in the asymptotic regime, further research on finite-blocklength codes is required on the path towards practical, secure communication systems. This work provides the first experimental characterization of a deep learning-based, finite-blocklength code construction for multi-tap fading wiretap channels without channel state information. In addition to the evaluation of the average probability of error and information leakage, we examine the designed codes in the presence of fading in terms of the equivocation rate and illustrate the influence of (i) the number of fading taps, (ii) differing variances of the fading coefficients, and (iii) the seed selection for the hash function-based security layer.

\end{abstract}

\section{Introduction}
\label{sec:intro}
Physical layer security (PLS) aims to guarantee security in communications system by directly integrating it into the physical layer. In contrast to conventional means of security that rely entirely on computational constraints of the adversary, PLS focuses solely on information-theoretic measures to describe provable notions of security~\cite{bloch2011}.

A fundamental model in PLS is the wiretap channel where the sender (Alice) wants to transmit a confidential message~$M\in\{0,1\}^k$ to the legitimate receiver (Bob). This message is encoded into a sequence~$X^n \in \mathcal{X}^n$ of length~$n$, that is sent over the channel and whose noisy version is received by Bob as~$Y^n\in \mathcal{Y}^n$. However, a malicious eavesdropper (Eve) could be able to learn information about this message via her channel observations~$Z^n\in \mathcal{Z}^n$.~\cite{wyner1975} and~\cite{csiszar1978} showed that the average probability of decoding error and the information leakage can be made arbitrarily small for infinitely long codewords, using a random-coding argument. Moreover, conventional channel codes, such as low-density parity check (LDPC)~\cite{thangaraj2007} and polar codes~\cite{mahdavifar2011}, have been adopted to construct secrecy capacity-achieving wiretap codes (WTC). In practical systems requiring low latency and hence, short packets, the assumption of infinite blocklength is no longer valid. Accordingly,~\cite{yang2019} derived achievability and converse bounds on the secrecy capacity in the non-asymptotic regime.

Modular coding schemes for the wiretap channel, composed of an error-correcting code and a security component were first proposed by~\cite{bellare2012}, bridging the gap between information-theoretic and cryptographic security.~\cite{wiese2021} and~\cite{torres-figueroa2021} applied this modular approach in the context of semantic security by deploying universal hash functions (UHF) as the security component. Furthermore,~\cite{rana2023} proposed the usage of deep learning-based channel codes as the reliability module in combination with an UHF and evaluated the performance of these codes for Gaussian wiretap channels. However, research on finite-blocklength wiretap codes for fading channels is still scarce. Recently,~\cite{mamaghani2024} provided a secrecy performance analysis of finite-blocklength transmissions in fading channels with instantaneous channel state information (CSI) of Bob's channel available at the transmitter.

In this work, we evaluate the seeded modular wiretap code design approach from~\cite{rana2023} on realistic channel models, namely multi-tap Rayleigh fading, and assume complete absence of CSI as a worst-realistic-case for secure communication. We first assess its performance in terms of the average probability of error and information leakage for a constant communication rate scenario. We further demonstrate how the fading channel can increase the equivocation rate of the system in comparison with the Additive White Gaussian Noise (AWGN) channel. Moreover, we illustrate the benefits of more channel taps as well as stochastical degradedness in terms of the information leakage. Finally, in contrast to schemes based on classical channel codes, we find that the selection of seeds for the UHF does not have an influence on the Hamming and Lee distances and, therefore, on the information leakage of the overall system. Our analysis provides fundamental insights into neural wiretap code designs for fading channels.

\section{System Model}
We consider a real-valued,~$T$-tap fading wiretap channel~$\left(\mathcal{X}, p_{YZ\mid X}, \mathcal{Y} \times \mathcal{Z}\right)$ given by
\begin{align}
    Y_i = \sum_{t=0}^{T-1}\left|H_{Y,t}\right| X_{i-t} + N_{Y,i}\text{,} \quad
    Z_i = \sum_{t=0}^{T-1}\left|H_{Z,t}\right| X_{i-t} + N_{Z,i}
    \label{eq:channel}
\end{align}
where~$N_{Y,i}$ and~$N_{Z,i}$ denote the zero-mean Gaussian random variables with variances~$\sigma_Y^2 = (2R_rE_b/N_{0,Y})^{-1}$ and~$\sigma_Z^2 = (2R_rE_b/N_{0,Z})^{-1}$. We define~$E_b/N_{0,Y}$ and~$E_b/N_{0,Z}$ as the per-bit energy to noise power spectral density ratio on Bob's and Eve's channel, respectively, and~$R_r$ as the encoding rate of the reliability layer (see Section~\ref{subsubsec:reliability_layer}). Due to this scaling we allow for a fair comparison of codes with different rates.~$\left|H_{Y,t}\right|$ and~$\left|H_{Z,t}\right|$ denote the magnitudes of the~$t$-th tap fading coefficients that follow Rayleigh distributions such that~$H_{Y,t}\sim\mathcal{CN}(0,\omega_{Y}^2 T^{-1})$ and~$H_{Z,t}\sim\mathcal{CN}(0,\omega_{Z}^2 T^{-1})$, which are normalized with respect to the number of fading taps to meet the unit power constraint. Note that choosing~$T>1$ results in intersymbol interference (ISI).

\begin{definition}[{\hspace{1sp}\cite{mittelbachSensingAssistedSecureCommunications2025}}]
    \label{def:stochastically_degraded_fading}
    The fading wiretap channel in~\eqref{eq:channel} is stochastically degraded if~$H_{Y}/{\sigma_Y^2}$ is stochastically larger than~$H_{Z}/{\sigma_Z^2}$, i.e., for all~$h\geq0$, we have
    \begin{align}
        \bar{F}_{H_Y} \left({h}/{\sigma_Y^2} \right) \geq \bar{F}_{H_Z} \left({h}/{\sigma_Z^2} \right)
    \label{eq:stochastic_degrad}
    \end{align}
    where~$\bar{F}_{X}(x) = \prob(X\geq x)$ is the complementary cumulative distribution function of a real-valued random variable~$X$.
\end{definition}



\subsection{Wiretap Coding}
Codes for the wiretap channel aim to provide a certain level of security in a scenario where the confidential communication over the channel is eavesdropped by an illegitimate user. In order to quantify their performance, one resorts to the following metrics:
\begin{itemize}[leftmargin=*]
    \item We measure the \emph{reliability} of the system in terms of the average probability of error at the legitimate receiver Bob, i.e.,~$P_e \estimates \prob\ [\hat{M} \neq M]$, where~$\hat{M}$ denotes the decoded message. Practically, it will be estimated by the block error rate (BLER), averaged over a campaign of Monte Carlo (MC) simulations.
    \item The \emph{secrecy} is determined by the amount of information about the source message~$M$ that is leaked to Eve via her channel observations~$Z^n$. This leakage metric is denoted by~$L \estimates I(M;Z^n)$.
\end{itemize}
\begin{definition}[\hspace{1sp}\cite{rana2023}]
An~$(n,k,P)$ code is~$\epsilon$-reliable if~$P_e \leq \epsilon$  and~$\delta$-secure if~$L \leq \delta$. Moreover, a secure communication rate~$R_s={k}/{n}$ is~$(\epsilon,\delta)$-achievable with power constraint~$P$ if there exists an~$\epsilon$-reliable and~$\delta$-secure~$(n,k,P)$ code.
\end{definition}

\begin{figure}[ht!]
\vspace{-0.2cm}
    \centering
    \includegraphics[width=.48\textwidth]{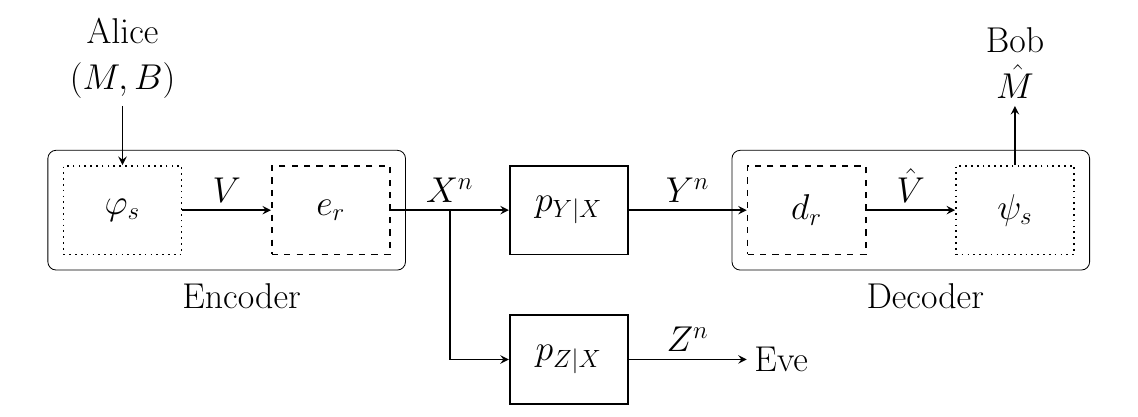}
    \caption{Modular wiretap code design consisting of the reliability layer~{$(e_r, d_r)$} and the security layer~{$(\varphi_s, \psi_s)$}.}
    \label{fig:system_model}
\end{figure}

We consider the modular coding scheme proposed in~\cite{rana2023}, depicted in~\autoref{fig:system_model}. This scheme guarantees the reliability and security constraints by an implementation consisting of two separately designed layers. A key advantage of this separable approach is its flexibility with respect to the redesign of any of these layers. In the following, we will briefly discuss the details of each layer.

\subsubsection{Reliability Layer}
\label{subsubsec:reliability_layer}
The encoder-decoder pair~$(e_r, d_r)$ of the reliability layer is implemented by an artificial neural network. Using an autoencoder structure~\cite{oshea2017}, the encoder and decoder parts are jointly optimized to minimize the BLER. Similar to classical channel codes, the encoder adds redundancy by learning new representations of the source messages in order to make them more robust against perturbations caused by the channel, while the decoder aims to recover the original message by exploiting the redundancy. \autoref{tab:autoencoder_config} shows the configuration of the autoencoder that is composed of a sequence of fully-connected (FC) layers with either rectified linear units (ReLU) or linear activation functions. It is trained using stochastic gradient descent (SGD) with the categorical cross-entropy loss function between the one-hot encoded message as ground truth and the softmax output of the decoder. This loss function inherently optimizes for the BLER~\cite{wiesmayr2023}. The encoding rate of the reliability layer is then given as $R_r=q/n$ where~$q$ is the length of the input sequence and~$n$ is the code's blocklength.


\begin{table}[t!]
\vspace{-0.2cm}
\caption{Configuration of the autoencoder network.}
\vspace{-0.2cm}
\label{tab:autoencoder_config}
\footnotesize
\begin{center}
\begin{tabular}{c c c c}
\toprule
&{Layers}&Input Size& Output Size \\
\midrule
Encoder & One-hot Encoder   & $q$   & $2^q$\\
        & FC Layer + ReLU   & $2^q$ & $2^q$\\
        & FC Layer + Linear & $2^q$   & $n$\\
        & Normalization     & $n$   & $n$\\
\midrule
Decoder & FC Layer + ReLU   & $n$   & $2^q$\\
        & FC Layer + Softmax& $2^q$   & $q$\\
\bottomrule
\end{tabular}
\end{center}
\vspace{-0.7cm}
\end{table}

\subsubsection{Security Layer}
The security layer aims to control the amount of information about the source message~$M$ that is leaked to Eve via her channel observations~$Z^n$. This can be achieved by the use of a 2-universal hash function (2-UHF)~$\psi_s$ and its inverse~$\varphi_s$, as defined below. Both of these functions rely on a seed~$s \in \mathcal{S}=\{0,1\}^q\setminus\{0\}^q$ that is assumed to be known by all parties. In addition, for every message~$M\in\{0,1\}^{k}$, Alice samples a uniformly distributed random bit sequence~$B\in\{0,1\}^{q-k}$ as a means of randomization to the output of~$\varphi_s$.
\begin{definition}[\hspace{1sp}\cite{hayashi2010}]
Let~$\mathcal{A}$ and~$\mathcal{B}$ be finite sets and~$\mathcal{F}$ a subset of the set of all mappings~$\mathcal{A} \rightarrow \mathcal{B}$, then~$\mathcal{F}$ is a family of 2-UHFs if~$\prob(F(x) = F(y)) \leq |\mathcal{B}|^{-1}$ where~$x,y \in \mathcal{A}$ and~$F$ is uniformly drawn from~$\mathcal{F}$.
\end{definition}

On the encoder side, we define the inverse function as
{%
\setlength{\belowdisplayskip}{0.1cm}%
\setlength{\abovedisplayskip}{0.1cm}%
\begin{align}
    \varphi_s : \{0,1\}^k \times \{0,1\}^{q-k} &\rightarrow \{0,1\}^q\text{,} \\
    (m,b) &\mapsto s^{-1} \odot (m||b)
\intertext{where~$\odot$ denotes the multiplication in~$\operatorname{GF}(2^q)$ and~$(\cdot||\cdot)$ the concatenation of two bit sequences. After passing through the reliability layer~$d_r$ within the decoder, the source message~$M$ is recovered by applying the hash function}
    \psi_s : \{0,1\}^q &\rightarrow \{0,1\}^k\text{,} \\
    v &\mapsto (s \odot v)_k
\end{align}%
}where~$(\cdot)_k$ denotes the truncation of the bit sequence to the~$k$ most significant bits.

\subsection{Mutual Information Estimation}
We want to estimate the information leakage in terms of the mutual information~$I(M;Z^n)$. As the analytical expressions for the respective joint and marginal probability distributions are unknown, we resort to a sample-based estimation approach. This can practically be accomplished by deploying the \emph{mutual information neural estimator} (MINE)~\cite{belghazi2018} that is based on a variational representation of the Kullback-Leibler divergence and provides a strongly consistent lower bound on the mutual information. We can parameterize a function~$T_\theta$, represented by a neural network, to obtain an estimate of the mutual information as
\vspace{-0.3cm}
\begin{align}
    \hat{I}(M;Z^n) \estimates \sup_{\theta\in\Theta} \Biggl[ & \frac{1}{N} \sum_{i=1}^N T_\theta (m(i),z^n(i)) \nonumber\\
    &- \log \left( \frac{1}{N} \sum_{i=1}^N e^{T_\theta (\tilde{m}(i),\tilde{z}^n(i))} \right) \Biggr]
\end{align}
where~$N$ input samples~$(m(i),z^n(i))$ are drawn from~$p_{MZ^n}$ and~$(\tilde{m}(i),\tilde{z}^n(i))$ from~$p_M p_{Z^n}$, respectively. Throughout this work, the parameterized neural network of MINE is implemented by a sequence of four FC layers with 400 neurons and ReLU activation each, except for the last layer that is followed by a linear activation.

MINE often exhibits high bias and variance in high-dimensional spaces~\cite{song2020}. Thus, as a precautionary measure, we consider a regime of dimensionality ($n\leq16$) that lies within a range in which MINE is tested to provide reliable estimates. Moreover, the size of our neural network is sufficiently large to be able to learn more complex probability distributions.

\vspace{-0.3cm}
\section{Main Results}
The following numerical simulations were conducted in PyTorch and accelerated by an NVIDIA RTX A5000 GPU. All autoencoder models were trained for a specific channel configuration between Alice and Bob. Moreover, they were evaluated for this channel configuration only, i.e., one model per scenario, as the scope of this work is rather on the neural coding schemes' reliability-security trade-off in the context of fading channels than on the networks' ability to generalize over a broad set of channels. The same applies to the MINE models, each of which aims to estimate the mutual information for a specific setting.

The training of the reliability layer was carried out using~$100$ epochs over a data set of~$10^6$ samples at~$E_b/N_{0,Y}=\SI{5}{\dB}$, which we consider as a fixed operating point for Bob. This step was performed without the UHF-based security layer. For the training of MINE, we spent~$20$ epochs over a set of~$2\cdot10^5$ samples. All models were trained using SGD with the Adam optimizer~\cite{kingmaAdamMethodStochastic2017} with a learning rate of~$0.001$ and a batch size of~$1000$. In addition, for MINE, we employed an exponentially decaying learning rate to ensure a smooth convergence. The full MC simulations deploying the trained models were executed with a samples size of~$10^6$.

\subsection{Average Error Probability and Information Leakage}
\label{subsec:const_rates}
\begin{figure*}[ht!]
\centering
    \begin{subfigure}[t]{0.32\textwidth}
    \centering
        \begin{tikzpicture}[trim axis left, trim axis right]
            \begin{semilogyaxis}[
                    legend style={nodes={scale=0.55, transform shape}, legend columns=2, column sep=0.05cm},
                    legend cell align={left},
                    legend pos=north west,
            		xmin=3, xmax=17,
            		ymin=1e-3, ymax=1e0,
            		grid=both,
            		xlabel={\footnotesize{$n$}},
            		ylabel={\footnotesize{$P_e$}},
            		grid style={dashed,gray!30},
                    scale=1.0,
                    width=1.0\textwidth,
            	]
                \addplot[color=plot0, very thick, mark=*, mark options={solid}] table [col sep=comma, x=blocklength, y=bler_source_msg] 
                    {plots/data/const_rates/awgn/leakage_bler_vs_blocklength_bob_5_eve_0.csv};
                \addplot[color=plot1, very thick, mark=*, mark options={solid}] table [col sep=comma, x=blocklength, y=bler_source_msg] 
                    {plots/data/const_rates/rayleigh-1-tap/leakage_bler_vs_blocklength_bob_5_eve_0.csv};
                \addplot[color=plot2, very thick, mark=*, mark options={solid}] table [col sep=comma, x=blocklength, y=bler_source_msg] 
                    {plots/data/const_rates/rayleigh-3-tap/leakage_bler_vs_blocklength_bob_5_eve_0.csv};
        
                \legend{
            		{\small{AWGN}},
                    {\small{1-Tap Rayleigh}},
                    {\small{3-Tap Rayleigh}}
            	}
                
            \end{semilogyaxis}
        \end{tikzpicture}
        \caption{Average error probability~$P_e$ at Bob over blocklength~$n$ for~$E_b/N_{0,Y}=\SI{5}{\dB}$.}
        \label{subfig:const_rates_bler_blocklength}
    \end{subfigure}
    ~
    \begin{subfigure}[t]{.32\textwidth}
        \centering
        \begin{tikzpicture}[trim axis left, trim axis right]
        
            \begin{semilogyaxis}[
        		legend style={nodes={scale=0.55, transform shape}},
                legend cell align={left},
                legend pos=south west,
        		xmin=3, xmax=17,
        		ymin=2e-3, ymax=5e0,
        		grid=both,
        		xlabel={\footnotesize{$n$}},
        		ylabel={\footnotesize{$L$ in bits}},
        		grid style={dashed,gray!30},
                scale=1.0,
                width=1.0\textwidth,
        	]
    
                \addplot[color=plot0, very thick, mark=*, mark options={solid}] table [col sep=comma, x=blocklength, y=mi_leakage_eve] {plots/data/const_rates/awgn/leakage_bler_vs_blocklength_bob_5_eve_0.csv};
                \addplot[color=plot1, very thick, mark=*, mark options={solid}] table [col sep=comma, x=blocklength, y=mi_leakage_eve] {plots/data/const_rates/rayleigh-1-tap/leakage_bler_vs_blocklength_bob_5_eve_0.csv};
                \addplot[color=plot2, very thick, mark=*, mark options={solid}] table [col sep=comma, x=blocklength, y=mi_leakage_eve] {plots/data/const_rates/rayleigh-3-tap/leakage_bler_vs_blocklength_bob_5_eve_0.csv};
                \addplot[color=plot0, very thick, mark=x, densely dashed, mark options={solid}] table [col sep=comma, x=blocklength, y=mi_leakage_eve] {plots/data/const_rates/awgn/leakage_bler_vs_blocklength_bob_5_eve_-5.csv};
                \addplot[color=plot1, very thick, mark=x, densely dashed, mark options={solid}] table [col sep=comma, x=blocklength, y=mi_leakage_eve] {plots/data/const_rates/rayleigh-1-tap/leakage_bler_vs_blocklength_bob_5_eve_-5.csv};
                \addplot[color=plot2, very thick, mark=x, densely dashed, mark options={solid}] table [col sep=comma, x=blocklength, y=mi_leakage_eve] {plots/data/const_rates/rayleigh-3-tap/leakage_bler_vs_blocklength_bob_5_eve_-5.csv};
            	

                \legend{
                    {\small{AWGN,~$\SI{0}{\dB}$}},
                    {\small{1-Tap Rayleigh,~$\SI{0}{\dB}$}},
                    {\small{3-Tap Rayleigh,~$\SI{0}{\dB}$}},
                    {\small{AWGN,~$\SI{-5}{\dB}$}},
                    {\small{1-Tap Rayleigh,~$\SI{-5}{\dB}$}},
                    {\small{3-Tap Rayleigh,~$\SI{-5}{\dB}$}}
            	}
        
            \end{semilogyaxis}
        
            
        \end{tikzpicture}
        \caption{Information leakage~$L$ to Eve over blocklength~$n$ for varying~$E_b/N_{0,Z}$.}
        \label{subfig:const_rates_leakage_blocklength}
    \end{subfigure}
    ~
    \begin{subfigure}[t]{.32\textwidth}
        \centering
        \begin{tikzpicture}[trim axis left, trim axis right]
            \begin{semilogyaxis}[
            		legend style={nodes={scale=0.55, transform shape}, legend columns=1, column sep=0.1cm},
                    legend cell align={left},
                    legend pos=south west,
            		xmin=0, xmax=9,
            		ymin=7e-2, ymax=4e0,
            		grid=both,
            		xlabel={\footnotesize{$T$}},
            		ylabel={\footnotesize{$L$ in bits}},
            		grid style={dashed,gray!30},
                    scale=1.0,
                    width=1.0\textwidth,
            	]
             
                \addplot[color=plot0, very thick, mark options={solid}] table [col sep=comma, x=num_taps, y=leakage_awgn] {plots/data/const_rates/num_taps/leakage_num_taps_eve_0_n_8.csv};
                \addplot[color=plot0, very thick, mark=x, mark options={solid}] table [col sep=comma, x=num_taps, y=leakage_rayleigh] {plots/data/const_rates/num_taps/leakage_num_taps_eve_0_n_8.csv};
                \addplot[color=plot1, very thick, densely dashed, mark options={solid}] table [col sep=comma, x=num_taps, y=leakage_awgn] {plots/data/const_rates/num_taps/leakage_num_taps_eve_0_n_12.csv};
                \addplot[color=plot1, very thick, mark=x, densely dashed, mark options={solid}] table [col sep=comma, x=num_taps, y=leakage_rayleigh] {plots/data/const_rates/num_taps/leakage_num_taps_eve_0_n_12.csv};
                \addplot[color=plot2, very thick, densely dotted, mark options={solid}] table [col sep=comma, x=num_taps, y=leakage_awgn] {plots/data/const_rates/num_taps/leakage_num_taps_eve_0_n_16.csv};
                \addplot[color=plot2, very thick, mark=x, densely dotted, mark options={solid}] table [col sep=comma, x=num_taps, y=leakage_rayleigh] {plots/data/const_rates/num_taps/leakage_num_taps_eve_0_n_16.csv};
            	
            	\legend{
                    {\small{AWGN,~$n=8$}},
                    {\small{Rayleigh,~$n=8$}},
                    {\small{AWGN,~$n=12$}},
                    {\small{Rayleigh,~$n=12$}},
                    {\small{AWGN,~$n=16$}},        
                    {\small{Rayleigh,~$n=16$}}
            	}
            
            \end{semilogyaxis}
        \end{tikzpicture}
        \caption{Information leakage~$L$ to Eve over number of fading channel taps~$T$ for~$E_b/N_{0,Z}=\SI{0}{\dB}$.}
        \label{subfig:const_rates_leakage_num_taps}
    \end{subfigure}
    \caption{Reliability and security evaluation of the designed WTC for constant rates~$R_s=1/4$, $R_r=1/2$,~$E_b/N_{0,Y} > E_b/N_{0,Z}$, and~${\omega_{Y}^2} = {\omega_{Z}^2} = 1$.}
    \label{fig:const_rates}
    \vspace{-0.5cm}
\end{figure*}

For the first characterization of the constructed WTC, we examine the previously defined reliability and security metrics. In order to provide a fair comparison among codes of different blocklengths, we keep the rates constant as~$R_s=1/4$ and~$R_r=1/2$. Moreover,~$E_b/N_{0,Y}= \SI{5}{\dB}$ is kept fixed for all simulations, while~$E_b/N_{0,Z} = \{-5, 0\}\,\si{\dB}$ is selected to investigate the scenario when Bob's advantage with respect to noise level declines. We first fix the distribution of fading coefficients for both channels by choosing~${\omega_{Y}^2} = {\omega_{Z}^2} = 1$.

\autoref{subfig:const_rates_bler_blocklength} depicts the average probability of error of Bob over varying blocklengths. It shows that the multi-tap fading channel poses a greater challenge to the decoder and, therefore, introduces more errors. However, Bob’s learned decoder takes advantage of the symbol correlations caused by ISI, as the BLER decreases in the case of 3-tap Rayleigh fading. We noticed a similar pattern in experiments with a comparable rate~$(7,4)$-Hamming code using maximum-likelihood sequence estimation.

Furthermore, the leakage to Eve over varying blocklengths is displayed in~\autoref{subfig:const_rates_leakage_blocklength} for two different noise levels. At~$E_b/N_{0,Z}=\SI{0}{\dB}$, the leakage for the fading channels is generally lower than for the AWGN channel. However, while the single-tap fading case maintains a constant gap to the AWGN case, the leakage for the 3-tap fading channel closes this gap for higher blocklengths. The reason for this observation could be an increase in the dependence among the channel observations~$Z^n$ resulting from the convolution of symbols. This effect is particularly significant for large~$n$, where the same~$3$-tap sliding window of channel coefficients uses several steps to move across the whole block. A similar observation is made at~$E_b/N_{0,Z}=\SI{-5}{\dB}$, where the leakage of the 3-tap channel even surpasses the AWGN case.

Moreover, we assess the leakage to Eve with an increasing number of channel taps, i.e., a higher amount of ISI within one block, illustrated in~\autoref{subfig:const_rates_leakage_num_taps}. While for the case of block fading, for~$T=1$, there is only a slight reduction in leakage with respect to the AWGN case, the leakage can be further lowered for channels with an increased number of fading taps because the amount of randomness introduced by the channel increases. The modest increase in leakage in the range~$T\in[2,4]$ before the start of the monotonic decay can again be explained with the higher dependence among the observations caused by ISI which becomes more pronounced with larger~$n$.

\subsection{Equivocation Rate}
In the context of the wiretap channel scenario, equivocation~$H(M|Z^n)$ quantifies Eve's uncertainty about the confidential message~$M$ upon observing~$Z^n$~\cite{bloch2011}, where~$H(\cdot)$ is the Shannon entropy. Normalized to blocklength~$n$, we can define the equivocation rate~$R_e$ as
\begin{align}
    R_e = \frac{H(M|Z^n)}{n} \leq \frac{H(M)}{n} = R_s
\end{align}
where the last equality follows from the assumption of uniformly distributed message bits such that~$H(M)=k$. The equality~$H(M|Z^n) = H(M)$ is achieved when~$M$ and~$Z^n$ are independent, i.e., complete uncertainty about message~$M$ given~$Z^n$. From the definition of mutual information, one can further express~$R_e$ in terms of the leakage~$I(M;Z^n)$ as
\begin{align}
    R_e = \frac{H(M) - I(M;Z^n)}{n} = R_s - \frac{I(M;Z^n)}{n}\text{.}
\end{align}
Using the same settings as in the previous scenarios, we display~$R_e$ based on the estimation for~$I(M;Z^n)$ in~\autoref{fig:equivocation_rate}. For the case~$E_b/N_{0,Z}=\SI{-5}{\dB}$, i.e., a~$\SI{10}{\dB}$ gap between Bob and Eve,~$R_e$ ranges above~$0.2$ bits/channel use. However, fading only yields a slight increase in the equivocation rate, as the channel noise is the factor that dominates the amount of information leakage. When the~$E_b/N_{0}$-wise gap between Bob and Eve decreases, i.e., the channel conditions for Eve improve,~$R_e$ is naturally lowered as more information is leaked. Nevertheless, in this case, the effects of the fading channels become more pronounced as~$R_e$ is increased by around~$0.035$ bits/channel use for all blocklengths. From an operational perspective, in scenarios when Bob's advantage over Eve with respect to channel noise is only minor, the fading characteristics can further boost secure communication rates. This finding aligns with various theoretical results on the benefits of fading for secure communications~\cite[Ch. 5.2]{bloch2011}. However, in contrast to our scenario, most of them assume some degree of CSI knowledge.
\begin{figure}[b!]
\vspace{-0.5cm}
    \centering
    \begin{tikzpicture}[trim axis left, trim axis right]
    \begin{axis}[
    		legend style={nodes={scale=0.55, transform shape}, legend columns=2, column sep=0.15cm},
            legend cell align={left},
            legend pos=south west,
    		xmin=3, xmax=17,
    		ymin=0, ymax=0.3,
            minor y tick num=5,
    		grid=both,
    		xlabel={\footnotesize{$n$}},
    		ylabel={\footnotesize{$R_e$ in bits/channel use}},
    		grid style={dashed,gray!30},
            scale=.75
    	]

        \addplot[color=plot0, very thick, mark=*, mark options={solid}] table [col sep=comma, x=blocklength, y=equivocation_rate] 
            {plots/data/equivocation_rate_leakage/awgn/equivocation_rate_per_blocklength_eve_0.csv};
        \addplot[color=plot1, very thick, mark=*, mark options={solid}] table [col sep=comma, x=blocklength, y=equivocation_rate] 
            {plots/data/equivocation_rate_leakage/rayleigh-1-tap/equivocation_rate_per_blocklength_eve_0.csv};
        \addplot[color=plot0, very thick, mark=x, densely dashed, mark options={solid}] table [col sep=comma, x=blocklength, y=equivocation_rate] 
            {plots/data/equivocation_rate_leakage/awgn/equivocation_rate_per_blocklength_eve_-5.csv};
        \addplot[color=plot1, very thick, mark=x, densely dashed, mark options={solid}] table [col sep=comma, x=blocklength, y=equivocation_rate] 
            {plots/data/equivocation_rate_leakage/rayleigh-1-tap/equivocation_rate_per_blocklength_eve_-5.csv};
        \addplot[mark=none, very thick, solid, black, samples=2] coordinates {(4,0.25) (16,0.25)};
    	
    	\legend{
            {\small{AWGN,~$\SI{0}{\dB}$}},
            {\small{1-Tap Rayleigh,~$\SI{0}{\dB}$}},
            {\small{AWGN,~$\SI{-5}{\dB}$}},
            {\small{1-Tap Rayleigh,~$\SI{-5}{\dB}$}},
            {\small{$R_s=k/n$}},
        }
    \end{axis}
    \end{tikzpicture}
    \caption{Equivocation rate~$R_e$ over blocklength~$n$ for varying~$E_b/N_{0,Z}$.}
    \label{fig:equivocation_rate}
\end{figure}

\subsection{Fading Coefficient Variance Analysis}
\label{subsec:stochastically_degraded_fading}
Next, we consider the scenario of Eve's channel being stochastically degraded with respect to Bob's only via the variances~$\omega_{Y}^2$ and~$\omega_{Z}^2$, which define the Rayleigh distribution of the fading coefficients, whereas the noise levels of both channels stay the same, i.e.~$\sigma_Y^2 = \sigma_Z^2$. In this case, condition~\eqref{eq:stochastic_degrad} is fulfilled if~$\omega^2_{Y}/\omega^2_{Z}\geq 1$. In the following experiment, we will keep~$\omega_Y^2=1$ fixed and only vary~$\omega_Z^2$. The communication rates~$R_s$ and~$R_r$ are chosen as in Section~\ref{subsec:const_rates}. In~\autoref{fig:stochastically_degraded_fading} we observe that lowering~$\omega_{Z}^2$, significantly reduces the leakage. This is expected since a Rayleigh distribution with a more narrow spread and its mode shifted closer to zero results in a higher number of lower-magnitude realizations of fading coefficients, which will ultimately weaken the average signal strength on Eve's channel.

\begin{figure}[b!]
\vspace{-0.5cm}
    \centering
    \begin{tikzpicture}[trim axis left, trim axis right]
    \begin{semilogyaxis}[
    		legend style={nodes={scale=0.55, transform shape}, legend columns=1, column sep=0.15cm},
            legend cell align={left},
            legend pos=south east,
    		xmin=3, xmax=17,
    		ymin=7e-2, ymax=4e0,
    		grid=both,
    		xlabel={\footnotesize{$n$}},
    		ylabel={\footnotesize{$L$ in bits}},
    		grid style={dashed,gray!30},
            scale=.75
    	]
     
        \addplot[color=plot0, very thick, mark=*, mark options={solid}] table [col sep=comma, x=blocklength, y=mi_leakage_eve] 
            {plots/data/const_rates/awgn/leakage_bler_vs_blocklength_bob_5_eve_0.csv};
        \addplot[color=plot1, very thick, mark=*, mark options={solid}] table [col sep=comma, x=blocklength, y=mi_leakage_eve] 
            {plots/data/const_rates/rayleigh-1-tap/leakage_bler_vs_blocklength_bob_5_eve_0.csv};
        \addplot[color=plot1, very thick, mark=x, densely dashed, mark options={solid}] table [col sep=comma, x=blocklength, y=mi_leakage_eve] 
            {plots/data/const_rates/degraded/rayleigh-1-tap/leakage_bler_vs_blocklength_bob_5_eve_0.csv};
        \addplot[color=plot2, very thick, mark=*, mark options={solid}] table [col sep=comma, x=blocklength, y=mi_leakage_eve] 
            {plots/data/const_rates/rayleigh-3-tap/leakage_bler_vs_blocklength_bob_5_eve_0.csv};
        \addplot[color=plot2, very thick, mark=x, densely dashed, mark options={solid}] table [col sep=comma, x=blocklength, y=mi_leakage_eve] 
            {plots/data/const_rates/degraded/rayleigh-3-tap/leakage_bler_vs_blocklength_bob_5_eve_0.csv};
    	
    	\legend{
            {\small{AWGN}},
            {\small{1-Tap Rayleigh,~$\omega_Z^2=1$}},
            {\small{1-Tap Rayleigh,~$\omega_Z^2=0.5$}},
            {\small{3-Tap Rayleigh,~$\omega_Z^2=1$}},
            {\small{3-Tap Rayleigh,~$\omega_Z^2=0.5$}},
    	}
    \end{semilogyaxis}
    \end{tikzpicture}
    \vspace{-0.1cm}
    \caption{Information leakage~$L$ to Eve over blocklength~$n$ for varying~$\omega_Z^2$ and~$E_b/N_{0,Y}=E_b/N_{0,Z}=\SI{0}{\dB}$.}
    \label{fig:stochastically_degraded_fading}
\end{figure}
\vspace{-0.1cm}
\subsection{Seed Selection}
Up to this point, we assumed the same fixed set of seeds~$s$ as in~\cite{rana2023}. However, the choice of these seeds and its effects on the randomization in the security layer of the system remain unclear. The authors of~\cite{frank2022} examined a modular wiretap code design involving a polar code-based reliability layer and a similar UHF-based security layer with respect to Eve's \emph{advantage}. This distinguishing security metric is characterized by the probability that Eve is able to distinguish between two confidential messages~$m_1$ and~$m_2$ given her observations~$Z^n$, maximized over all possible sets~$(m_1,m_2)$. They found the seed selection to be crucial for this metric. Specifically, they identified two classes of seeds based on the dispersion of the distribution of the average Hamming distance~$d_H$ between the encoding results of~$m_1$ and~$m_2$ with a string of random bits~$b$, using the same seed~$s$. As the reliability layer~$e_r$ in our system comprises both channel coding and modulation, we will apply an~$l$-level quantization stage to the real-valued outputs of the encoder to enable a similar examination. The quantized encoder output is denoted by~$\bar{e}_r$. In order to provide a more nuanced distance evaluation for these non-binary words, we will consider the Lee distance~$d_L$ over the~$l$-ary alphabet~\cite{lee1958} in addition to the Hamming distance, defined as
\vspace{-0.2cm}
\begin{align}
    d_L(u,v) = \sum_{i=1}^n \min\left(\left| u_{i} - v_{i} \right|, l -\left|u_{i} - v_{i}\right|\right)
\end{align}
where~$u,v\in\{0,l-1\}^n$ denote encoded and quantized blocks.

For the fixed configuration~$k=4$,~$q=8$, and~$n=16$, we calculated the distances~$d_H((\bar{e}_r(\varphi_{s}(m_1,b)),\bar{e}_r(\varphi_{s}(m_2,b))$ and~$d_L(\bar{e}_r(\varphi_{s}(m_1,b),\bar{e}_r(\varphi_{s}(m_2,b))$ for each seed~$s$, including all possible combinations of messages~$m_1$ and~$m_2$ with~$m_1 \neq m_2$ and random bits~$b$. For the quantizer, we chose a step size of~$l=16$. As an example,~\autoref{fig:seed_dispersion} depicts the empirical distribution for both distance metrics for~${s=(0,0,0,0,0,0,1,1)}$. In contrast to the findings of~\cite{frank2022}, it was observed that these statistics remain the same for all possible seeds, i.e., the selection of seeds does not have a significant influence on the output of the designed WTC. Moreover, we verified this observation by the information leakage analysis that does not show major deviations when choosing different seeds.

\begin{figure}[t!]
\vspace{-0.25cm}
\centering
    \begin{subfigure}[t]{.23\textwidth}
    \centering
        \begin{tikzpicture}[trim axis left, trim axis right]
            \begin{axis}[
                    ybar,
            		xmin=-1, xmax=7,
            		ymin=0,
            		grid=both,
            		xlabel={\footnotesize{$d_H$}},
            		ylabel={\footnotesize{Number of message pairs}},
            		grid style={dashed,gray!30},
                    scale=0.42,
                    xtick=data,
                    bar width=0.2cm,
                    xtick={0, 2, 4, 6},
                    ytick={2e+3,4e+3},
                    scaled y ticks=base 10:-3,
                    xtick pos=left,
                    ytick pos=left
            	]
                \addplot[fill, color=plot0] table [x=bins,y=counts, col sep=comma] {plots/data/seeds/hamming_seed_3.csv}; 
                
            \end{axis}
        \end{tikzpicture}
    \end{subfigure}
    ~
    \begin{subfigure}[t]{.23\textwidth}
    \centering
        \begin{tikzpicture}[trim axis left, trim axis right]
            \begin{axis}[
                    ybar,
            		xmin=15, xmax=55,
            		ymin=0,
            		grid=both,
            		xlabel={\footnotesize{$d_L$}},
            		ylabel={\footnotesize{Number of message pairs}},
            		grid style={dashed,gray!30},
                    scale=0.42,
                    xtick=data,
                    bar width=0.01cm,
                    xtick={20, 30, 40, 50},
                    scaled y ticks=base 10:-3,
                    xtick pos=left,
                    ytick pos=left
            	]
                \addplot[fill, color=plot0] table [x=bins,y=counts, col sep=comma] {plots/data/seeds/lee_seed_3.csv}; 
                
            \end{axis}
        \end{tikzpicture}
    \end{subfigure}
    \vspace{-0.1cm}
    \caption{Histograms of Hamming distances~$d_H$ and Lee distances~$d_L$ for the~$16$-step quantized encoder output for all possible combinations of message pairs~$(m_1, m_2)$ with~$m_1\neq m_2$ and random bits~$b$, for the WTC with~$k=4$,~$q=8$ and~$n=16$ and~${s=(0,0,0,0,0,0,1,1)}$.}
    \label{fig:seed_dispersion}
    \vspace{-0.6cm}
\end{figure}
\vspace{-0.1cm}
\section{Conclusion}
We adopted a framework for a modular neural wiretap code design consisting of a learned channel code as reliability layer and a UHF as security layer for multi-tap fading channels without CSI, and experimentally assessed its performance. In comparison to the AWGN case, we showed that while sacrificing reliability, the security of the system will benefit from the presence of fading, as the leakage to Eve is significantly lowered. Specifically, in terms of the equivocation rate, the fading characteristics can make a crucial difference in scenarios with only a small noise-level advantage of the legitimate party. Moreover, the increase of the number of fading taps as well as a lower variance of the fading coefficients of Eve's channel can further reduce the information leakage. Finally, we found that, in contrast to modular wiretap codes designs that involve classical codes, the choice of seeds does not make a significant difference for our autoencoder-based setup. 

Future work will focus on extensions towards larger, more practical blocklengths. In this context, a more powerful technique of MI estimation should be introduced. Moreover, as the current implementation of the reliability layer relies on the basic autoencoder setup, we inherit certain limitations with respect to the scalability due to the one-hot representation of source messages. Suitable candidates may be found in concatenated approaches of hybrid channel codes\cite{gunluConcatenatedClassicNeural2023a}.
\vspace{-0.1cm}

\bibliographystyle{IEEEtran}
\bibliography{ieee/IEEEabrv,refs}

\end{document}